\date{}
\documentclass[preprint]{revtex4}
\usepackage{epsf}
\begin{document}

\title
{Finite size effects and localization properties of disordered quantum wires 
with chiral symmetry}

\author{G. Chiappe\footnote[2]{Member of CONICET, Argentina} and  M. J. S\'anchez\footnote[2]{Member
 of CONICET, Argentina}}
\vspace{1cm}

\affiliation{ Departamento de F\'{\i}sica  J. J. Giambiagi,
Facultad de Ciencias Exactas  y Naturales,  \\
Universidad de Buenos Aires.
Ciudad Universitaria, 1428 Buenos Aires, Argentina.}

\begin{abstract}
Finite size effects in the localization properties of disordered quantum wires are analyzed 
through conductance calculations.
Disorder is induced by introducing vacancies at random positions in the wire and 
thus preserving the chiral symmetry.

For quasi one-dimensional geometries and low concentration of vacancies, an exponential 
decay of the mean conductance with the wire 
length is obtained  even at the center of the energy band. For wide wires,  finite size 
effects cause the conductance to decay following a non-pure exponential law. We propose  an 
 analytical formula  for the mean conductance that reproduces accurately the numerical 
data for both geometries.
However, when  the concentration of vacancies  
increases above a critical value, a transition towards the suppression of the conductance 
occurs. This is a signature of the presence of ultra-localized states trapped in 
finite regions of the sample.

\end{abstract}

\maketitle

\vspace{2cm}
\section{Introduction}

Disorder effects in the transport properties of quantum wires have been the subject
of extensive investigations in recent years. Among these works,  models with pure
off-diagonal disorder have been studied in connection to  peculiar properties that seem
to differentiate them from models with pure diagonal disorder only 
\cite{eilmes,brouwer1,brouwer2,mudry,verges}.
Unusual phenomena like  anomalies in the density of states (DOS) and in the localization 
properties of quantum wires have been reported  for a  
random hopping model of disorder in the vicinity of energy $E=0$. 
As an example, in Refs. \cite{brouwer1,brouwer2,mudry}  the authors
pointed out  that, for a wire with an odd number of channels and at the band center, the 
conductance decays algebraically with the wire length  rather than 
exponentially, as is usual in the problem of Anderson localization \cite{lee}.
 These behaviors have been attributed  to  the existence of a delocalized state at $E=0$ 
that arises as a consequence of an additional lattice symmetry, absent in models with pure 
diagonal disorder. Due to this symmetry, referred as chiral, the eigenvalues
appear in pairs $\pm E$ and the spectrum is symmetric with respect to $E=0$
\cite{brouwer1,brouwer2,mudry}.  

However, for other models of disorder with chiral symmetry the existence of a delocalized
state near the center of the band is still a subject of debate  \cite{miller,furu}. 
One source of discrepancy is the large localization
length obtained near the center of the band, that makes difficult to decide from 
numerical data whether or not states are localized.

In view of this controversy,  we find instructive to study  an 
alternative model of disorder that, like the random hopping model, has  chiral symmetry. 
The localization properties will be inferred from  conductance 
calculations  performed in wire like geometries.
Disorder is induced introducing  a number of vacancies  $N_v$ at random positions in a 
regular lattice that defines the wire. Vacancies represent sites with an infinite 
energy that block the motion of the electrons.
 Therefore the only way an electron 
may propagate across the sample is through  a path with no vacancies.
This model could  be thought as a 
limiting case of the random site percolation problem \cite{koslo} known
as the binary-alloy model \cite{kir,souk}. 

For quasi-one dimensional wires  we show that, unlike the random hopping model, the present 
 model of disorder  exhibits exponential localization  at the band center irrespective 
of the parity in the number of transmission channels. We propose  an analytical formula   
that reproduces the behavior of the mean conductance as a function of the vacancy 
concentration. For wide wires, as finite size effects  become important in the 
mesoscopic regime, a detailed  analysis of the influence of the geometry  on the conductance 
and on the localization properties will  be performed.

As the vacancy concentration increases above a critical value we  show that 
ultra-localized states are formed. These states influence dramatically the value of the 
conductance and a transition towards the suppression of the conductance in the wires is 
observed.

The present model supports the conjecture that the chiral symmetry appears to  be a necessary 
but not a sufficient condition for the existence of anomalies at the band center in the 
localization properties of disordered quantum wires.

\section{Conductance calculations}

The wires are stripes  of length $L $ and  width $W $ defined on a square lattice. 
We employ the tight-binding  Hamiltonian 
with a single atomic level per site :
\begin{equation}
\label{tb}
H = - \sum_{i,j_{i}} {\hat{c}_{i}}^{\dagger} {\hat c}_{j_{i}} \; ,
\end{equation}
where the operator  $\hat{c_i}$ destroys an electron on  the site $i$.
 All the hopping integrals are taken equal to $-1$ and restricted to nearest neighbors.
$j_{i}$ labels {\it only} the existing nearest-neighbors of site $i$ after
introducing $N_v = \alpha \; W \; L $ vacancies at random positions.
Ideal leads of width $W$ are attached
at both ends of the sample through  hopping integrals equals to $-1$ and enter in
the Hamiltonian  Eq.(\ref{tb}) as complex self-energies \cite{datta}. 
The Landauer formalism \cite{landau} is employed
to calculate the conductance  for each  realization
of disorder. Open boundary conditions are imposed in the direction transverse to
conduction.

\begin{figure}
\epsfysize=15cm
\epsffile {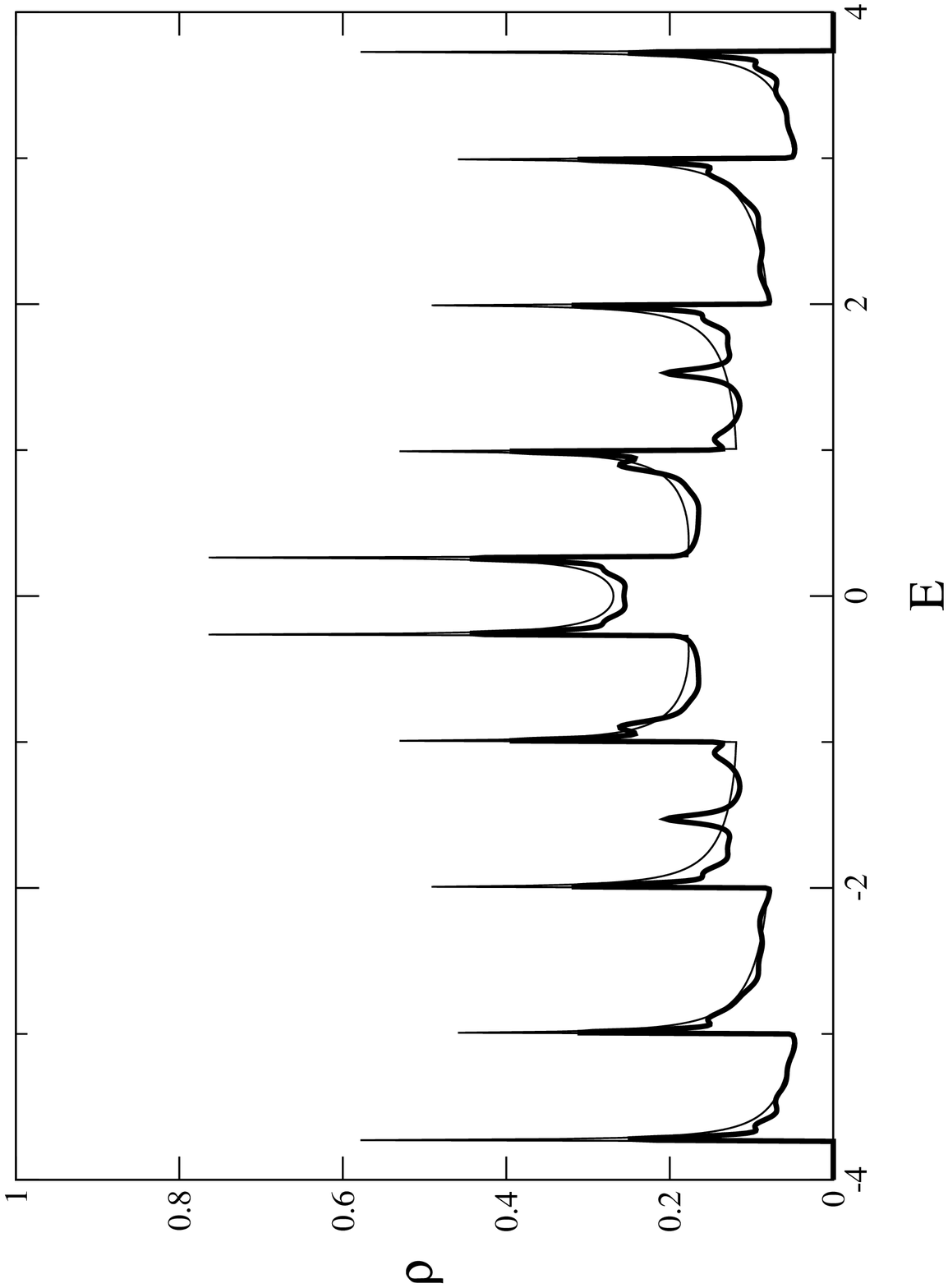}
\caption {Density of states (DOS) $\rho$ as a function of the energy $E$ 
for a clean quantum wire (thin line) of $W=5$ and $L= 100$ and for 
the same geometry but when $\alpha = .02$ (thick line). 
The chirality manifests in the symmetry with respect to the 
$E=0$ axis.}
\label{1}
\end{figure}

The present model of disorder has chiral (particle - hole) symmetry. This is 
illustrated in Fig. \ref{1}, where we
plot  the DOS as a function of the energy $E$
for a wire of  $W = 5 $ and $L= 100$. For comparison we show the ideal situation 
(without vacancies) and when the concentration of vacancies is $\alpha = .02$.
In both cases, the   chiral symmetry is  manifested in 
the symmetry of the spectrum with respect to the $E=0$ energy axis.

The conductance of a quantum wire of width $W$ which supports  $N(E)$ open  
channels, 
\begin{equation}
G \equiv \sum_{n=1}^{N(E)} g^{(n)} \; = \frac{e^2}{\hbar} \sum_{n,m=1}^{N(E)} 
|t_{n,m}|^2 \; ,
\end{equation}
is related to the transmission amplitudes $t_{n,m}$ connecting the incoming
mode $n$ at the entrance  lead with the outgoing mode $m$ in the exit 
lead. Therefore for a perfect  wire  with no vacancies it is well known that the 
dimensionless conductance  $g \equiv G/ ({e^2}/{\hbar})$ increases in one unit  when a 
new incoming channel is opened.
For $N(E)$ open channels is $g  = N(E)$, and at the band center 
$g(E= 0) = N(E=0) = W$.

In the following  we will  perform conductance calculations as a function of the vacancy 
concentration $\alpha$. We will focus  on the results
at the band center ($E = 0 $). This is the special
energy value for which anomalies  in the conductance have been reported in the 
random hopping model of disorder \cite{brouwer1,brouwer2,mudry}. 
We will  obtain an 
analytical formula that reproduces quite satisfactory the behavior of the mean conductance
as a function of the vacancy concentration for quasi-one dimensional wires and also for 
wide wires in which $W \lesssim L$.

\subsection{Quasi-one dimensional wires}
 
Here we present   conductance calculations for quasi-one dimensional wires where
$W << L$. Averages over configurations of disorder have been  performed to obtain the mean
conductance, that unless other specification we will denote for simplicity with the symbol $\bf g$.
The total number of vacancies for each concentration  is $N_v = \alpha \;  L \;  W $.
 
We start analyzing the effect of $N_v = 1$ vacancy on the conductance.
Due to the fact that the perfect wire has translation symmetry along the longitudinal direction
($\bf x$), the wire with one vacancy has inversion symmetry along  this direction with respect to
a transverse line that contains the vacancy.
Therefore the wave functions of the system with one vacancy can be written as   linear combinations
of products of plane waves along the direction of conduction and  functions with a node in the
site of the vacancy in the 
the transverse direction to conduction (along the $\bf y$ line  that
contains the site of the vacancy). Then, due to  the chiral symmetry,  the number of active 
transverse modes at energy $E=0$ is determined by the number of independent  solutions of 
 the  isolated  vertical line that contains  $W-1$ sites.  It is exactly $W-1$.
Therefore    the value of the 
conductance is reduced in one unit with respect to the value in the clean wire. That is
 ${\bf g}_{1}  = W - 1$, where the subscript refers to the number
of vacancies  and it is understood that $E = 0$. 

If a second vacancy is added  at the same $y$ than the
first one but at a different coordinate $x$, the number of transverse modes remains the same
as  in the case of one vacancy. Therefore the conductance  at $E=0$ is still $W-1$. On the 
other hand, if the second vacancy is put 
at the same $x$  coordinate than  the first one, but  at a different $y$ coordinate, the 
number of transverse channels is reduced by two with respect to the value in the clean case. 
Therefore the conductance at $E=0$ would  be $W-2$.
For any other spatial location of the second vacancy, the separability among the transverse 
and longitudinal modes is lost. This implies that the conductance  will not  
be $\it {exactly}$ an integer number any more. 
In order to have  an insight about how  the conductance is modified in a generic situation, 
we will keep in mind that  the effect of a single  
vacancy in a clean wire is   to cut exactly one channel regardless  the spatial location 
of the vacancy in the sample.
  
Alternatively, we  could calculate the  (mean)conductance of a wire with only
 one vacancy   as the  sum   of the individual conductances of  $W$ single channel 
conductors, each one with  a {\bf mean conductance } of $ (W - 1)/ W $. 
This gives ${ g}_{1} = W  \left( ( W - 1) / W \right) = W - 1$, as we previously obtained. 
In the following  to compute the conductance we imagine the  wire  as composed of  
single-channel  conductors in parallel. This approximation  neglects 
interference effects among the different
transverse channels when vacancies are added and relies in the fact that the wire 
is quasi-one dimensional ({\it i.e.} the energy separation between transverse modes is 
very large).
The effect of the second   vacancy will be to suppress 
exactly one channel in  a `` clean wire" composed by W single channel conductors, each one 
with   a  {\bf mean conductance } of $ (W - 1)/ W $.
Therefore the conductance for $N_v= 2$  can be calculated as the sum of the conductances of 
$W-1$ single 
channel conductors each one with an average conductance of  $(W - 1 )/ W $. This gives 
${ g}_{2}   = (W-1) \;  (W - 1)/ W  \; = W \; \left((W - 1) / W \right)^2 $. 

Following this simple scheme   up to $N_v$ vacancies we obtain

\begin{equation}
\label{gmedia1}
{ g}_{N_{v}}  = W \; \left( \frac{W - 1}{ W} \right)^{N_v} = \;
W \exp \left( - \;  \alpha  \; W \log \; \left( {\frac{W}{W-1}}\right)  L \right) \; ;
\end{equation}
where we have replaced in the right hand side $N_{v} =\alpha  \; W  \; L $ in order to show 
explicitly the exponential decay  of the mean conductance  with the length of the sample.

Therefore, from Eq.(\ref{gmedia1})  one concludes that the quasi-one dimensional wire with 
$N_v$ vacancies  could be model as  $W$ single channel conductors in parallel, each one 
with an  average conductance of $\left((W - 1) / W \right)^{N_v} $.

\begin{figure}
\epsfysize=15cm
\epsffile {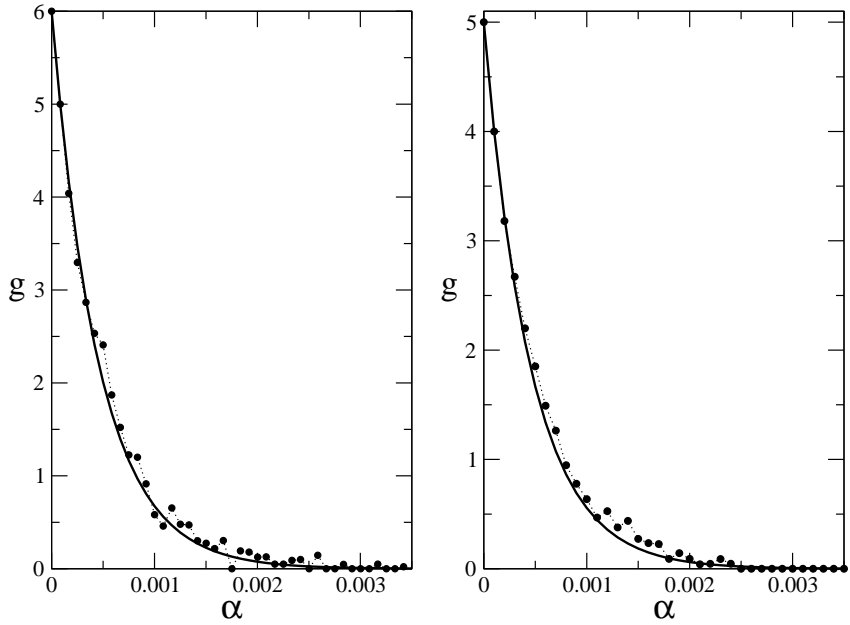}
\caption {Left panel: Mean conductance $\bf g$ (for $100$ realizations of 
disorder) at the band center $E=0$, as a function 
of the vacancy concentration $\alpha$ for a quantum wire of $W=6$ and 
$L=2000$. The dots are  the numerical results and
the full line is the  analytical estimate given by 
Eq.~(\ref{gmedia1}). Right panel: Idem as left panel but for a wire of $W=5$ and 
$L=2000$.}
\label{2}
\end{figure}

In the left (right) panel of Fig.~\ref{2} we show  $\bf g$ as a function of the vacancy 
concentration $\alpha$ for a wire of  width $ W = 6 \; ( W= 5)  $ and $L = 2000$. 
The solid line  in both panels shows the exponential law, Eq.(\ref{gmedia1}),
 that follows accurately the  numerical results (dots) for an ample range of values 
of $\alpha$. 
It is interesting to remark that the localization properties of this model 
do not depend on the parity of the  number of transverse channels as it happened in the random
hopping model studied in Refs. \cite{brouwer1,brouwer2,mudry}.

An analytical estimate of the localization length $\xi$ can be obtained from 
Eq.(\ref{gmedia1}),
\begin{equation}
\label{lw}
\xi = \frac{1}{  \alpha  \; W \log \left( \frac{W}{W-1} \right)},
\end{equation}
 which in the limit of $W \sim 1$ gives $\xi \sim 0$. This is consistent with  the fact 
that only one vacancy suffices to suppress the conductance of  a one-dimensional wire,
that is  ${ g}_{1} = 0 $ for  $W = 1$, as we have previously obtained. 
Moreover, when $\alpha \sim 1/L $, that is when $ N_v \sim W$,  the conductance becomes
 lower than one, in agreement  with Eq.(\ref{gmedia1}) and the model of W 
independent channels.

In the limit $1 << W << L$ is
$\xi \sim 1 /  \alpha$. When  $W << L$,  it is satisfied that the mean free path 
${\ell}$ and the localization length $\xi$ are related by  
$\xi = \; W {\ell}$ \cite{datta}.  Thus ${\ell} = L/ N_v$, as expected 
for quasi-one dimensional samples.

\subsection{Wide wires: finite size effects}

When the aspect ratio  of the sample $r \equiv W /L $  increases interference effects 
between the transverse channels become relevant.
Therefore the conductance can not be calculated as a sum of the  conductances of 
one-dimensional wires. As a consequence one
should expect that Eq. (\ref{gmedia1})  departs from  the numerical results when the
aspect ratio $r \lesssim 1 $.

As was already mentioned, when vacancies  are added  at 
arbitrary locations in the sample, the wave functions can not be written as  products of
 functions of the  longitudinal coordinate by functions of the transverse coordinate. 
Disorder mixes the  transverse and longitudinal modes and then  the separability is lost.  

When the first vacancy is added, exactly one channel is suppressed. This is the 
maximum suppression possible   when a site is eliminated and it is due to the fact that 
the ``one vacancy problem'' is  separable,  as was already discussed.
However, due to the  mixing  between channels,  when a  new vacancy is added 
that maximum value is not attained and less than one channel is effectively eliminated. 
Therefore  the efficiency of the vacancies for decreasing the conductance is reduced. 

In a classical picture the carriers can  
now propagate  following non rectilinear paths
along the  direction of conduction, avoiding the obstacles (vacancies).  
Therefore, even  after including the rescaling of the mean conductance per 
channel each time a new vacancy is added (as it was  done   when we obtained
 Eq.(\ref{gmedia1})), the actual conductance for a given value of $\alpha$ should be
greater than the value predicted by the pure exponential decay Eq. (\ref{gmedia1}).

Taking into account the effect described above, we propose for wide wires 
the following  ansatz for the mean conductance as a function of $\alpha$,

\begin{equation}
\label{gmedia2}
{ \bf  g}  = W \;  \exp \left( - \;  \alpha  \; W \; \log{ \left( {\frac{W - \; \beta }{W- \; 1 + 
\beta}}\right)} \; L \right) \; ;
\end{equation}

with $\beta  \equiv (\alpha)^{r} \; ( 1 - \alpha ^{r/2})$. For $\alpha << 1 $ or 
$r <<1 $, $\beta  \to 0$ and the pure exponential decay predicted
in  Eq.(\ref{gmedia1}) is reobtained.

\begin{figure}
\epsfysize=15cm
\epsffile {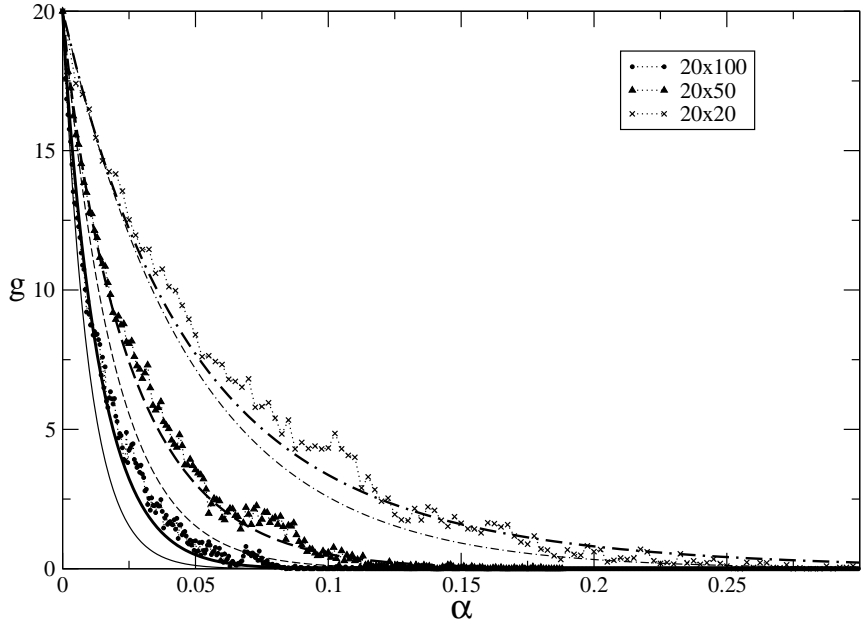}
\caption {Mean conductance $\bf g$  at  $E=0$, as a function 
of the vacancy concentration $\alpha$ for quantum wires of $W=20$ and 
$L=100$ (the dots are the numerical results,
the  full line is the  analytical estimate given by 
Eq.~(\ref{gmedia1}) and the thick full line is the  analytical estimate given by 
Eq.~(\ref{gmedia2})), $L= 50$ (the triangles are the numerical results,
the thin dashed line is the  analytical estimate given by 
Eq.~(\ref{gmedia1}) and the thick dashed line is the  analytical estimate given by 
Eq.~(\ref{gmedia2})) and $L= 20$ (the crosses are the numerical results,
the thin dotted-dashed line is the  analytical estimate given by 
Eq.~(\ref{gmedia1}) and the thick dotted-dashed line is the  analytical estimate given by 
Eq.~(\ref{gmedia2})).}
\label{3}
\end{figure}

In Fig. \ref{3} we show the conductances
as a function of  $\alpha$  for three different   samples with $W= 20$ and $L=20, 50$ and $100$.
The aspect ratios are $ r = 1, 0.4$ and $ 0.2$ respectively.
It is clear that even for relatively small concentrations, the  pure exponential formula 
Eq.(\ref{gmedia1}) departs from  the numerical results. 
On the other hand Eq.(\ref{gmedia2}) gives a  quite satisfactory fit to the numerical results
even for large concentrations  (see the caption in Fig.\ref{3}).
 
\begin{figure}
\epsfysize=15cm
\epsffile{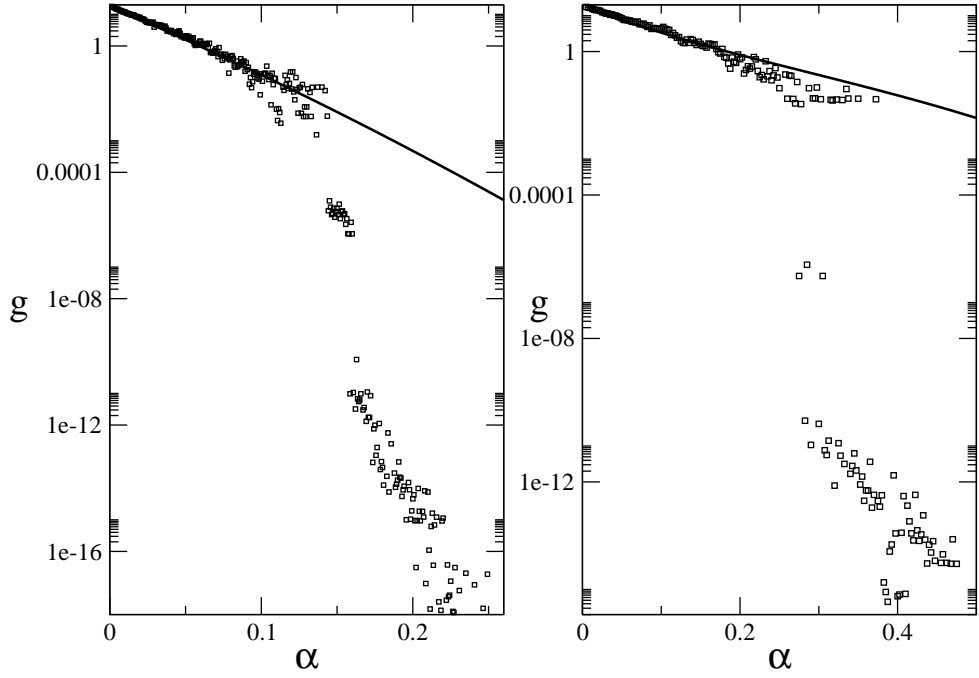}
\caption {Left panel: Log-Lin plot of the mean conductance $\bf g$  at  $E=0$ as a 
function of $\alpha$ for a quantum wire of $W=20$ and 
$L=70$ (the squares are the numerical results
and the  full line is the  analytical estimate given by 
Eq.~(\ref{gmedia2})). Right panel: Idem  as left panel but for a  wire of $W=20$ and 
$L=20$. }
\label{4}
\end{figure}

However, there is a critical 
value of the concentration  above which the conductance falls abruptly towards $0$. Fig.\ref{4}
shows in a log-lin plot the numerical values for the mean conductance together with those
predicted by Eq.(\ref{gmedia2}) for two samples with $W= 20$ and $L= 70$ and $L = 20$ 
respectively.
In both plots, and in spite of the numerical fluctuations, the fall of the conductance 
towards $0$ is clearly observed.
The suppression of the conductance is a signature of the presence  of ultra-localized states
formed  by  effect of the vacancies at $E= 0$. 
These ultra-localized states correspond to eigenstates of the disorder isolated $W \dot L $ 
stripe which persist even after the system is embedded with the leads 
({\it i.e.} these eigenstates extend along a distance $d < L$).
To illustrate this point, we evaluate the DOS for a  wire (stripe with leads)  
and  two  values of the concentration $\alpha$, below and above the critical value  
($\alpha_0$) respectively. 
In order to put in evidence the existence of ultra-localized states, we include
in the computation of the DOS a small imaginary part $\delta$, such that
$\rho(E) \equiv  Im \sum_{j}  1 / ( H - \; E + i\;  \delta )_{jj} $.
The small quantity $\delta$ does not modify the DOS when  ultra-localized states are 
absent (remember that we are computing the DOS for an open system). On the other hand, 
when the ultra-localized states exist, a single pole in the DOS will appear at $E= 0$. 
It corresponds to a peak with a height that increases linearly with $ 1/ \delta$ 
as $\delta \rightarrow 0$. 

\begin{figure}
\epsfysize=15cm
\epsffile{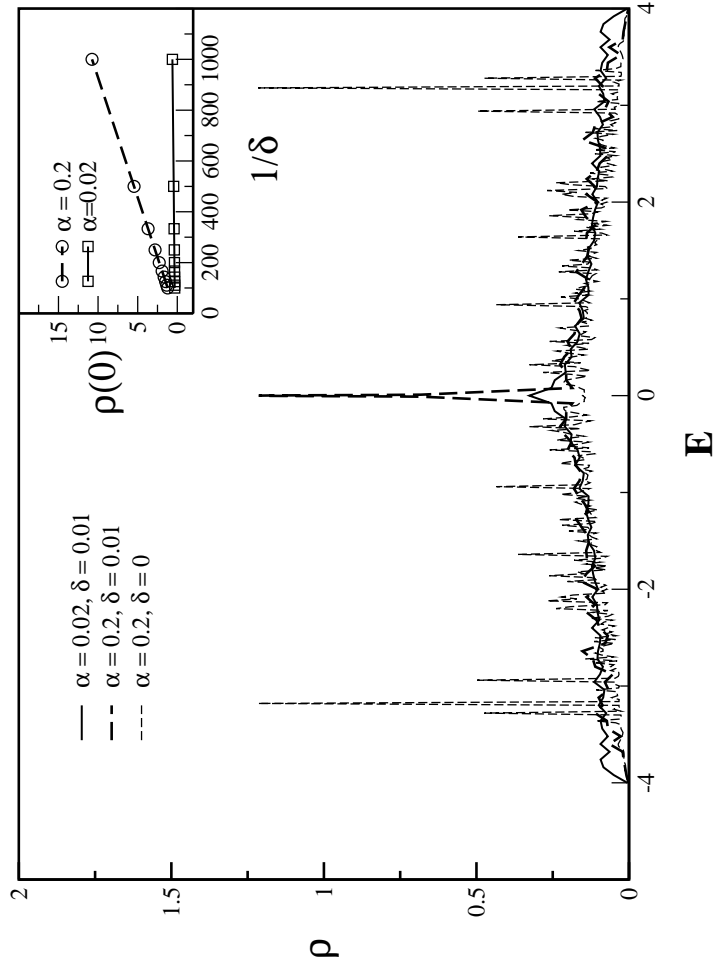}
\caption {DOS $\rho(E)$ for a wire of $W=20$ and 
$L=50$. The solid (dashed) line corresponds to a  concentration $\alpha = 0.02 \; (0.2)$ 
below (above) the critical concentration  $\alpha_0$. 
The peak at $E=0$ for $\alpha = 0.2$, $\rho(0)$, is a signature of the presence of 
localized states. In order to show that the peak represents  a Dirac's delta
function,  the inset shows that $\rho(0)$ increases linearly with 
$ 1/ \delta$ for $\alpha = 0.2$ whereas for $\alpha = 0.02$  it does not depend 
on $\delta$ ($\delta$ is the 
small imaginary part  defined to compute $\rho(E)$).}
\label{5}
\end{figure}

Figure \ref{5} shows the DOS as a function of the energy $E$, $\rho(E)$,
 for a wire of $L=50$ and $W=20$ and two  concentrations $\alpha= 0.02 < \alpha_0$ 
 and $\alpha= 0.2 > \alpha_0$ ($\alpha_0 \sim 0.12$ for this sample).

The  peak in the DOS at $E=0$ for  $\alpha= 0.2$ is the signature of the
presence of ultra-localized states which are  trapped in 
finite regions of the  sample  defined by the  spatial distribution of the 
vacancies.  It should be remarked that it is not necessary that the vacancies 
enclose completely the region in which the state is trapped, as it was already 
noted in Ref. \cite{kir}. 

It is worth to mention that the smooth part of the DOS (that is obtained
with $\delta = 0$) at $E=0$ has a dip but it is not zero, showing the coexistence of
 ultra-localized states and generic states at the same energy. 
As  the conductance falls down to zero for concentrations $\alpha > 
\alpha_0$, they should be spatially
disconnected (indeed the extended states must necessary come from the leads). 
Therefore when 
ultra-localized states are formed, the transport across the wire is forbidden.

As the aspect ratio of a sample  increases, the 
value of $\alpha_0$ around which the transition takes place also increases. 
This is consistent with 
the fact that  interference effects between  longitudinal and transverse channels 
(which are maximized for the square geometries) favors the conduction along the sample. 

For $\alpha < \alpha_0$, that is before the transition towards the suppression of 
the conductance occurs,  the non-pure exponential law Eq.(\ref{gmedia2}) implies that, 
unlike the quasi one-dimensional wires, it is not possible
 to define a localization length in the usual way ({\it i.e.} as the inverse of the 
decay rate in the exponential  law for the conductance as a function of the length $L$).

\section{Conclusions}

In this work we study the localization properties of  disorder quantum  wires.
Disorder is simulated distributing  vacancies at random positions in the sample and thus 
preserving the chiral symmetry.
Conductance calculations at $E=0$ have  been performed as a function of the 
concentration of vacancies $\alpha$, in quasi-one dimensional  and in  
wide wires, respectively.

For quasi-one dimensional wires  we have shown that the  conductance decays with 
the length of the wire following an exponential law irrespective of  the parity
in the number of transverse channels. This  is at odd with previous results 
reported for the random hopping model of disorder in which the scaling of the 
conductance at the band center would depend on the parity of the number of 
transverse channels \cite{mudry}.

We have  derived an analytical formula Eq.(\ref{gmedia1}) that 
reproduces quite accurately the behavior of the mean conductance for an ample range of 
concentrations $\alpha$.
 
For wide wires and low values of $\alpha$ a pure exponential decay of the mean 
conductance with the length $L$ has been  also observed. This exponential 
localization is consistent with the  quantum percolation transition   predicted for  
the binary-alloy model in 2d lattices at $\alpha = 0$ \cite{souk} (all the states are 
exponentially localized for any amount of disorder in a 2d lattice).
Therefore, in an infinite
system, only one vacancy is enough to suppress the conductance.

However for wide wires and greater values of $\alpha$, we found that 
the conductance decays following a non pure exponential law. We have proposed an 
ansatz, Eq.(\ref{gmedia2}),
that takes into account how  interference effects between  channels affect the conductance 
of the sample as the concentration $\alpha$ increases.
If  from Eq.(\ref{gmedia2}) we define 
\begin{equation}
\label{lw2}
\bar \xi  \equiv \frac{1}{  \alpha  \; W \log \left( \frac{W -\beta}{W-1 + \beta} \right)},
\end{equation}
this quantity  could be interpreted  as  a localization length
that depends on  the finite dimensions of the wire.

Finally, a new abrupt transition to the suppression of the conductance
 is observed when the concentration of vacancies $\alpha$
 goes beyond some critical  value $\alpha_0$ (which is indeed sample dependent) in a 
finite system. 
This transition, which is a signature of the formation at $E=0$ 
of ultra-localized states trapped in finite regions of the sample,  can 
 be also interpreted  as the quantum analog of the classical cluster percolation transition
 (CPT). Classically, the CPT and the 
suppression of the conductivity occurs at the same critical concentration \cite{stau}.
In a quantum system they could be  separated due to the fact that ultra-localized states
could be formed in an infinite cluster for a finite value $\alpha_0 > 0$.

The ultra-localized  states are poles at $E=0$ of the DOS, even after the leads 
are included, and therefore should be spatially separated  from any other
extended state of the leads. We should mention that ultra-localized states at 
$E=0$ have been already reported  for the binary  alloy model in isolated samples  
by Kirkpatrick and Eggarter \cite{kir}. 
In addition, we do not observe in the DOS a signature of the presence of ultra-localized 
states out of the band center when  the leads are included.

In an infinite  2d quantum system the 
conductance is suppressed as soon as a single vacancy is introduced, so the conductance
 is not a relevant magnitude to detect the {\bf quantum} CPT.
However, as in a finite quantum system  there is not a complete suppression of the
conductance  for finite values of $\alpha$,  the
conductance can be employed  to estimate the  critical concentration for 
the {\bf quantum} CPT. We are currently working along this line.

\acknowledgments
We would like to thank P. W. Brouwer and J. A. Verg\'es for useful discussions.

This work was partially supported by UBACYT (EX210) and CONICET.

\newpage

\newpage

\newpage

\newpage

\newpage

\end{document}